# Improving Network Performance with Affinity based Mobility Model in Opportunistic Network

Suvadip Batabyal, Parama Bhaumik

Dept. of Information Technology
Jadavpur University, Kolkata (WB), India
{mailto.sbatabyal@gmail.com, parama@it.jusl.ac.in}

*ABSTRACT*

*Opportunistic network is a type of Delay Tolerant Network which is characterized by intermittent connectivity amongst the nodes and communication largely depends upon the mobility of the participating nodes. The network being highly dynamic, traditional MANET protocols cannot be applied and the nodes must adhere to store-carry-forward mechanism. Nodes do not have the information about the network topology, number of participating nodes and the location of the destination node. Hence, message transfer reliability largely depends upon the mobility pattern of the nodes. In this paper we have tried to find the impact of RWP (Random Waypoint) mobility on packet delivery ratio. We estimate mobility factors like number of node encounters, contact duration(link time) and inter-contact time which in turn depends upon parameters like playfield area (total network area), number of nodes, node velocity, bit-rate and RF range of the nodes. We also propose a restricted form of RWP mobility model, called the affinity based mobility model. The network scenario consists of a source and a destination node that are located at two extreme corners of the square playfield (to keep a maximum distance between them) and exchange data packets with the aid of mobile 'helper' nodes. The source node and the destination node are static. The mobile nodes only help in relaying the message. We prove how affinity based mobility model helps in augmenting the network reliability thereby increasing the message delivery ratio and reduce message delivery latency.*

*KEYWORDS*

*OpNet, Random Waypoint, contact duration, inter-contact time, affinity based movement model, satellite node, degree of bias.*

## 1. INTRODUCTION

Opportunistic networks (shortly known as the OpNets) are a special class of Mobile Ad-hoc Networks where the network remains disconnected for most of the time and a complete path from source to destination does not exist. The nodes must leverage sporadic, sometimes unpredictable encounter events between nodes (i.e., when two nodes move within wireless communication range) to exchange control and data messages, enabling network wide communication. Hence, performance of an opportunistic network largely depends on the mobility of the nodes. A number of mobility models like RandomWayPoint, RandomDirection, MapBased Movement and many more exist and hence, the study and analysis of these mobility models becomes an essential task in the research of opportunistic networks. In the RWP movement model, the node density is highest at the centre of the playfield [1, 2] and probability of finding a node decreases as one moves away from the centre towards the border. This makes the border regions inaccessible since the probability of finding a node at the border is zero. As a result, a source node near the border remains disconnected most of the times. Hence, if we can make some nodes frequently available to the source, it can utilize this opportunity to *quickly* handover the packets to some nearby node. Based on this notion we propose a restricted form of RWP model called the *affinity based mobility model*. In this scheme we introduce some bias in

International Journal of Wireless & Mobile Networks (IJWMN) Vol. 4, No. 2, April 2012

node distribution, that is, there is a certain area where a node will spend most of the time or is most likely to be found. These specified areas may be points of high interest (hot-spot areas) where the nodes will move most frequently. We introduce this model in the subsequent section.

Message delivery probability largely depends on the degree of connectivity of the network. Degree of connectivity defines *how frequently* the mobile nodes encounter each other which again depend on parameters like area of playfield, RF range of the nodes, node velocity and spatial location of the nodes. In this paper we try to calculate the expected number of encounters, the link duration and the average inter-contact time between a mobile node and a static node located near the border of the playfield as packet delivery largely depends on these parameters. We relate how expected number of encounter decrease as the playfield area increases. Then we derive the desired inter-contact time and how performance largely depends on these parameters. Next we briefly study the generic RWP mobility model and explain how performance can be improved with affinity based mobility model.

Opportunistic network seems to be a promising future network technology and has already been successfully implemented in many projects like wildlife tracking (ZEBRANET [3], SWIM [4]), developing remote area communication (DAKNET [5]), interplanetary communication [6, 7], disaster management [8] and VANETs (Vehicular ad-hoc networks [9, 10]). The Vehicle-Assisted Data Delivery (VADD) protocol [11] in VANET uses a "carry-and-forward" strategy to allow packets to be carried by vehicles in sparse networks for eventual forwarding when another appropriate node enters the broadcast range, thereby allowing packets to be forwarded by relay in case of sparse networks. The aim of our paper is to facilitate wireless communication between two distantly located static nodes, with the help of a few mobile nodes without any pre-established infrastructure. We also try to minimize resource utilization, ensure maximum packet delivery and reduce packet delivery latency.

The paper is organized as follows. Section II highlights the related works with respect to mobility models in opportunistic networks and elicits challenges that still exist in this field. Section III highlights the factors affecting the performance of opportunistic network and numerical estimates of these factors. Section IV describes the proposed scheme. Section V illustrates the simulation results and inferences obtained from them. Section VI puts a concluding remark with a scope of future work.

## 2. RELATED WORKS

In literature, the terms Delay Tolerant Network and Opportunistic networks have been used interchangeably. Based on the DTN architecture defined in [12], we can say that DTNs exploits some knowledge of network topology which is lacking in opportunistic networks. Opportunistic networks are characterized by sparse connectivity, forwarding through mobility and fault tolerance. In such networks, transmissions are brought about by the moving nodes and hence, they adhere to *store-carry-forward* method. For this a new bundle layer protocol (Fig. 1) has been proposed in [2]. Since opportunistic networks are characterized by spontaneous message transmission brought about mobile nodes, studying mobility patterns has been an important part of the research. Figure 3 shows a classification of mobility models. Amongst all mobility models, the RWP mobility model has been studied to a great extent due to its simplicity and easy implementation and because of its highly stochastic movement characteristics. Though RWP mobility model has been widely studied and has been accepted as the standard, there are certain drawbacks associated with it. The spatial node distribution in a RWP mobility model is transformed from uniform to non-uniform distribution (fig. 2) with progress of time [13]. It finally reaches a steady state condition where the node density is maximum at the centre of the playfield, whereas the node density tends to zero as we move towards the border region. Moreover the average number of neighbours for a particular node periodically fluctuates over time. This phenomenon is called the density wave phenomenon [14] of the Random Waypoint model.

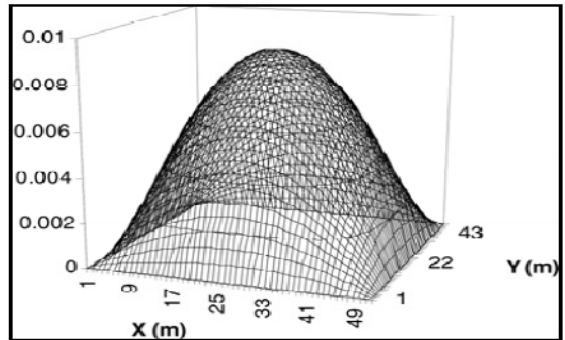

Figure 1. The OpNet protocols stack    Figure 2. Spatial node distribution in a square area

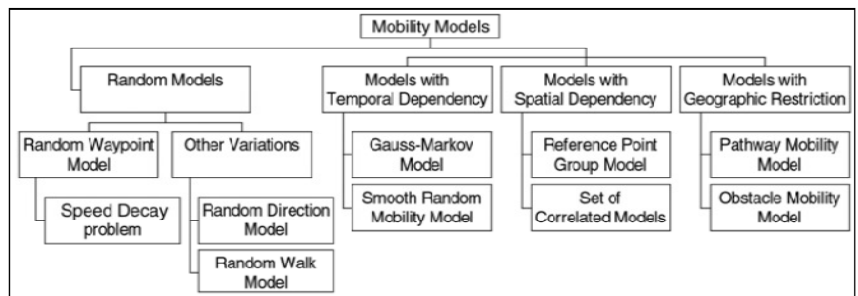

Figure 3. Categories of mobility models in MANET

## 3. MOBILITY PARAMETER ESTIMATION

The Random Waypoint Model was first proposed by Johnson and Maltz [15]. Soon, it became a 'benchmark' mobility model to evaluate the MANET routing protocols, because of its simplicity and highly stochastic movement characteristics. According to the theory of random process, **the Random Waypoint process has mean-ergodic property.** But RWP has several drawbacks such as non-uniform node distribution, speed decay problem and density wave phenomenon as pointed out in section II. In this section we study what are the factors that affect network throughput, overhead and packet delivery latency. We also highlight the various parameters that can be regulated to optimize these network parameters.

### 3.1. Underlying Scenario

We have considered square simulation field of size a by a (denoted by $A_{sim}$) with two static nodes viz. the source and the destination placed at two extreme corners of the playfield. We have considered a square field simulation area for the sake of simplicity although a circular or any arbitrary field shape may be considered. Our aim is to study the network performance and throughput of the opnet when the distance between the source and the destination is maximum. Only the two static nodes can generate data packets with the help of a message generating application (here we have simulated a ping-pong application). All the other nodes, known as the helper nodes, are mobile and move with a constant velocity with a pause time between intermediate transitions. They only help in relaying packets from source to destination. We assume that all nodes have same movement velocity (except for the source and the destination), RF range, bit rate, and buffer size (Table I).

### 3.2. Number of Encounters

Consider a static node $n_i$ having RF range 'R' meters placed at coordinates $(x_i, y_i)$ inside a square field of area $A_{sim}$ ($a \times a$). A helper node $n_j$ performs generic RWP movement with velocity v mts/sec. According to RWP mobility model the spatial node distribution in a square field of area $A_{sim}$ can be expressed as [15]:

$$f_{X,Y}(x,y) \approx \frac{36}{a^6}\left(x^2 - \frac{a^2}{4}\right)\left(y^2 - \frac{a^2}{4}\right), for\ x\ \in \left[-\frac{a}{2},\frac{a}{2}\right] and\ y \in [-\frac{a}{2},\frac{a}{2}]$$

A mobile node comes within the range of the static node when it passes through the bounded region $D(x_i - R \leq x \leq x_i + R, y_i - R \leq y \leq y_i + R )$

Hence, the probability that the mobile node can be found within the given coordinates is given by $p = P(x, y \in D) = \iint_D f_{X,Y}(x,y)\,dx\,dy \quad \ldots (i)$

The expected transition length of a mobile node within a square area of $A_{sim}$, as derived in [16] is $E[l] = .5214a$. Therefore a node having a constant velocity v mts/sec, the expected transition time between two random points is $E[T_l] = \frac{E[l]}{v}\ sec$. We denote this time as $T_{epoch}$. A node performing RWP randomly chooses a new point after each $T_{epoch}$. However, the transition time may vary depending upon the transition length. If the total simulation time is $T_{sim}$, we can say that the expected number of random points chosen (by a single mobile node) is $\frac{T_{sim}}{T_{epoch}}$. And such points may either lie inside $D$ or the node may cross $D$ to reach the destination with probability $p$ (as in (i)). Let the expected number of encounters (with both the source and the destination node) be denoted by $E[N_{encounter}]$. We plot the graph of simulation area against $N_{encounter}$ for three different node velocities. We have found that expected number of encounters (keeping all other factors constant) varies inversely as the cube of the simulation area (fig. 4).

$$E[N_{encounter}] = \frac{c}{A_{sim}^3} \quad \ldots (ii)$$

where 'c' is a constant factor. [For the proof refer Appendix A]

We can see that number of encounters decay as we increase the field area. Hence, number of encounters depends on the field area to a great extent. The constant 'c' in equation (ii) depends on factors like node velocity (v), RF range of the node (R) and $p$ which again depends upon the spatial location and the area of $D$. We know that $p$ is maximum for a node placed in the centre of the simulation field i.e. at coordinates (0, 0). Hence, a node placed in the middle of the square field is likely to have more number of encounters with the mobile node(s) rather than a node placed near the boundary.

### 3.3. Contact Duration and Inter-contact Time

Contact duration (or the link duration) and inter-contact time significantly affects message delivery reliability. Contact duration is the time a node stays within the RF range of another node i.e. the link duration between two nodes. Contact duration determines the amount of bytes transferred from one node to another (both uplink and downlink). RF range and node velocity determines the contact duration between the two nodes. With respect to the proposed scheme when one node is static and is located near the field boundary the expected contact duration is given by $E[T_{contact}] = \frac{2R}{v} \ \ldots (iii)$

Inter-contact time is the time elapsed between two successive contacts. Packets are dropped on TTL expiry or due to limited buffer size and thus greater the inter-contact time greater is the chance of packet being dropped. Number of encounters is an indirect measurement of inter-contact time i.e. as the number of encounters increase, the average inter-contact time decreases. It is often that the rate of message created at source overwhelms the buffer size and packets loss becomes inevitable. Hence, we must derive the desired inter-contact time so that less number of packets are dropped due to TTL expiry before they are delivered to the destination and hence reduce overhead.

Consider nodes having buffer size B bytes and rate of message created at source be λ bytes/sec. Hence, the inter-contact time should be $\Delta T \leq \frac{B}{\lambda}$. This is a necessary condition but not sufficient condition to prevent packet loss. Mobile nodes are often resource constraint devices and hence have a limited buffer capacity. Packet loss at source also depends on the byte transfer rate (bit-rate) of the communication link. Let the nodes have a bit-rate of b bytes/sec. Hence, packets are dropped if $N_1 > N_2$ where $N_1$: number of packets created during ΔT and $N_2$: number of packets dispersed/transferred during contact time. To reduce packet loss, $\lambda \times \Delta T \leq \frac{2Rb}{v}$. Hence,

$$\Delta T \leq \frac{2Rb}{\lambda v} \quad ....(iv)$$

We can infer that inter-contact time largely depends upon $T_{epoch}$ which again depends on the area of the playfield and the node velocity. It is also inversely proportional to $p$ which suggests that inter-contact time also depends on the spatial location of the node(s). Figure 5 shows the plot of area against link duration.

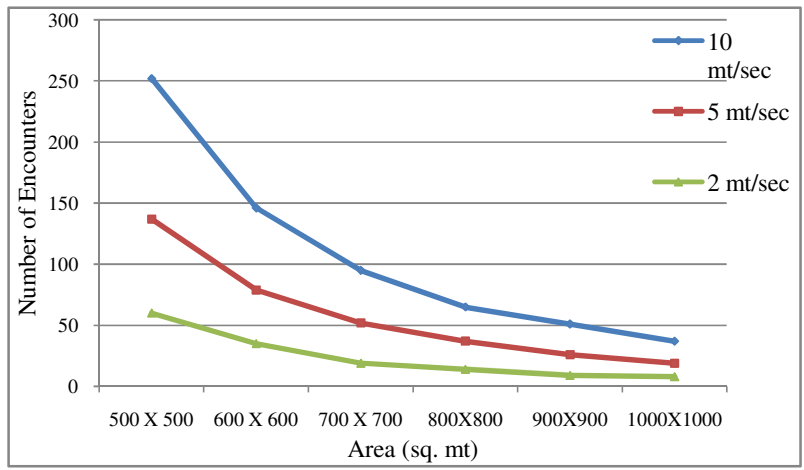

Figure 4. Graph- Area (sq. mt) against Number of encounters (sec)

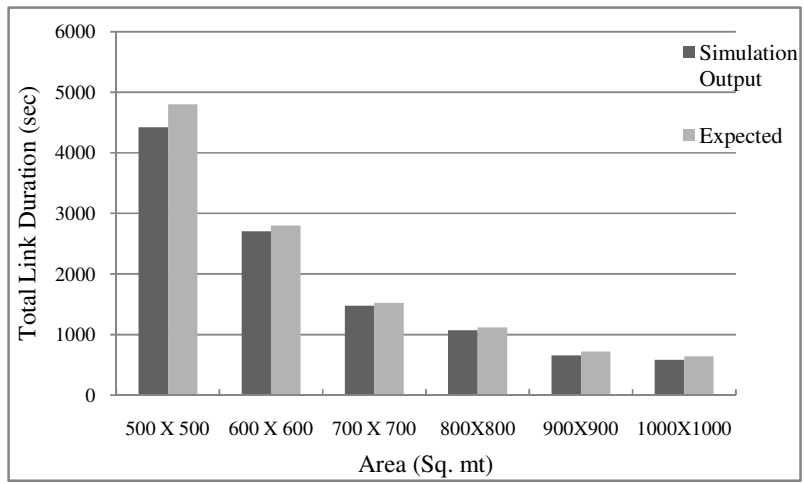

Figure 5. Graph- Area (sq.mt) against Total link duration (sec)

We can surely reduce the amount of packet loss at source by reducing the inter-contact duration. Inter-contact time can be reduced by: a) increasing the node density b) increasing the node velocity c) biasing some helper nodes (use of satellite nodes). Increasing the node density by increasing the number of helper nodes is a costly measure. Even node velocity cannot be increased beyond a certain limit. So the best possible way to reduce the inter-contact time is by biasing the helper nodes i.e. by using satellite nodes. For this we propose the affinity based mobility model as described below.

## 4. AFFINITY BASED MOBILITY MODEL

Since we have placed the source and the destination at two extreme corners of the square field, helper nodes rarely come in contact with the source and the destination. We have also shown that the expected number of encounters ($E[N_{encounter}]$) decrease as we increase the area. Hence, to increase the frequency of contact, we need to *bias* some of the helper nodes towards these regions so that the number of encounters with the source and the destination increases. That is, some of the helper nodes have affinity towards the source while some are more inclined to the destination (fig. 6). We call such biased helper nodes as the *satellite nodes*. Simulation results prove that affinity based mobility model helps in performance improvement with respect to message delivery probability and delivery latency.

The satellite nodes have degree of bias which suggests how frequently a node will visit a given area to which it is biased. The degree of bias is a value $0 \leq d \leq 1$. We call these biased nodes as the *satellite nodes* because their movement pattern is predisposed to the regions around the source/destination node. A value of $d \geq 0.5$ suggests a positive bias while a value having $d < 0.5$ suggests a negative bias. Let us calculate how often a node visits a given area with a given degree of bias.

Consider a square area 5000mts X 5000mts and a static source located at co-ordinates x=100, y=100 inside the square area having a predefined radio range. Another mobile node performs RWP movement. We know that in RWP movement model, a node *randomly* selects its next destination point, after a brief pause at the current location. For this we select a random variable $r(\xi)$ with normal distribution taking values $0 \leq r \leq 1$ with µ = 0.5. Hence, all the nodes are most likely to be found near the centre of the playfield as demonstrated in [17].

Now, let us bias the node movement by degree $d = 0.8$, i.e.

$$0 \leq x_{next} \leq 200, 0 \leq y_{next} \leq 200, \ if \ r \leq 0.8$$

$$200 < x_{next} \leq 5000, \ 200 < y_{next} \leq 5000, \quad if \ r > 0.8$$

Since $r(\xi)$ is a random variable having normal distribution, the percentage of area under the curve for $0 \leq r \leq 0.8$ is about 72.5%. Hence, the chance of finding the node inside the given area is about 72.5%. However, it should be noted that this bias does not mean that a given node will remain confined to the biased area only. It only enhances the number of hits or encounters with the source node with respect to the other nodes.

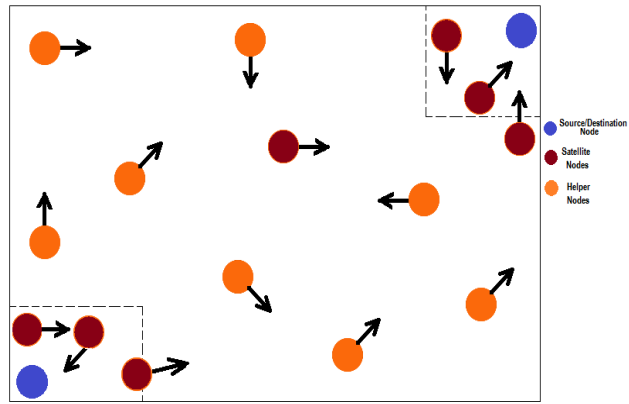

Figure 6. Schematic of a biased RWP showing the source/destination, helper and satellite nodes

## 5. SIMULATION RESULTS

We have used ONE [18, 19] (Opportunistic Network Environment) as our simulation tool and have conducted simulation based on the number of input parameters. The proposed affinity based mobility model was tested using the following routing algorithms: a) epidemic [20] b) binary spray-and-wait [21] c) prophet [22]. The following parameters were taken into consideration for the assessment of the proposed scheme:

a. Message delivery probability is defined as the fraction of messages having distinct message/packet id that are delivered. Packets are dropped once the TTL value expires.

$$Message\ delivery\ probability = \frac{No.\ of\ message\ delivered}{No.\ of\ message\ created}$$

b. In flooding, more than one copies of a message are relayed to ensure packet delivery with maximum probability. More the number of relayed packets greater is the overhead ratio.

$$Overhead\ ratio = \frac{message\ realyed - message\ delivered}{message\ delivered}$$

c. Message delivery latency is defined as the time taken by a message to reach the destination from the source. As the number of hops increase, message latency increases. We measure the *average* latency i.e. average of all the latencies of all nodes in the network.

We have used a square playfield of area 5000 X 5000 with the following simulation inputs:

TABLE I. GENERAL SIMULATION INPUT PARAMETERS

| | |
|---|---|
| Source node coordinates(x, y) | 500, 4500 |
| Destination node coordinates(x, y) | 4500, 500 |
| Node RF range | 80 meters |
| Bit rate | 2 Mbps |
| Packet Size | 1kB |
| Message TTL | 200 sec |
| Queue Size | 512kB |
| Node Velocity | 10mts/sec or 36kmph |
| Node stop time (min/max) | 5/10 sec |
| Message generation rate | 1 packet per 500 sec |
| Simulation time | 30000 sec |
| Number of nodes | 30 |

TABLE II. SIMULATION INPUT PARAMETERS FOR BIASED NODES

| | |
|---|---|
| Degree of bias | 0.8 |
| Source satellite nodes ($x_{min}$, $y_{min}$) | 0, 4000 |
| Source satellite nodes ($x_{max}$, $y_{max}$) | 1000, 5000 |
| Destination satellite nodes ($x_{min}$, $y_{min}$) | 4000, 0 |
| Destination satellite nodes ($x_{max}$, $y_{max}$) | 5000, 1000 |

For the simulation we have considered eight satellite nodes; four satellite nodes for the source and four for the destination. The remaining 20 nodes are unbiased helper nodes and perform generic RWP movement. Keeping all the simulation parameters (like node density, node velocity, etc) fixed, we have found that above configuration yields optimized results. Increasing the number of satellite nodes beyond four results in increased overhead without an increase in throughput. Again, lesser number of satellite nodes lowers throughput. The biased nodes have specified regions to which they have a greater affinity of movement. They are referred by coordinates ($x_{min}$, $y_{min}$ and $x_{max}$, $y_{max}$). Refer table 2 for details.

For binary spray-and wait we have considered initial number of message copies to be 6. For prophet router, we assumed predictability initialization constant, $p_0 = 0.75$, predictability transitivity scaling constant, $β = 0.25$ and predictability aging constant, $α = 0.98$ [Refer Appendix B & C for details].

### 5.1. Impact on message delivery probability

Figure 7 shows the performance of affinity based movement model with respect to packet delivery probability. We know that epidemic router performs best when delivery success rate is concerned and results show that success rate goes even higher with biased node movement. The packet delivery success rate is higher even for spray-and-wait and prophet routers. Message delivery probability is higher in affinity based movement model because messages created at the source spend less time in source queue and hence there are less chances of packet being dropped due to TTL expiry before it is delivered. As the satellite nodes frequently encounter the source node, packets spend less time at source. From the graph we can see the probability of message delivery is 0.775 for epidemic router with biased helper nodes while it is only 0.623 for generic RWP model.

### 5.2. Impact on overhead ratio

We know that spray-and-wait performs best when overhead ratio is concerned. Figure 8 shows that affinity based movement model has similar overhead ratio as compared to generic RWP model. Overhead ratio largely depends on the type of data dissemination technique used and is independent of the movement model considered. Hence, the proposed scheme has insignificant effect on the overhead ratio.

### 5.3. Impact on average latency

Affinity based mobility model has a positive impact on the average latency. Figure 9 show that this model reduces packet latency drastically in all the routing schemes. Average latency reduces by 41%, 44% and 26% for epidemic, spray-and-wait and prophet routers respectively. Packets spend less time waiting in the source queue and are quickly handed over to the satellite nodes. Similarly, at the destination side packets spend less time in the satellite queue and are instantly transferred to the destination node. Hence, a message spends less time in the network and hop quickly from source to destination.

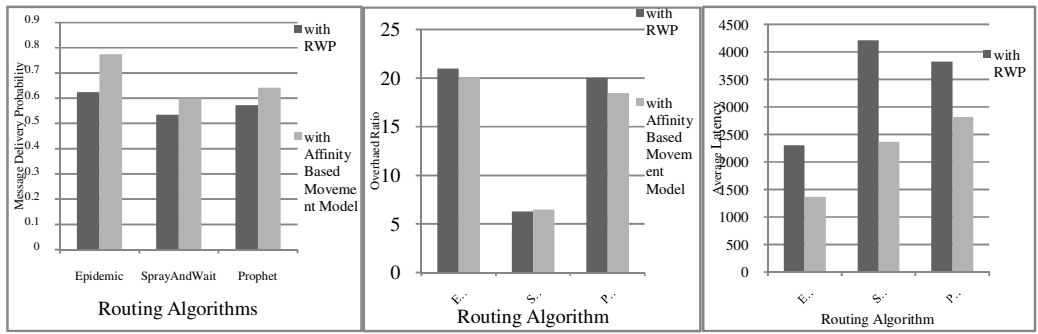

Figure 7. Graph- Comparison showing Message Delivery Probabilities

Figure 8. Graph - Comparison showing Overhead Ratios

Figure 9. Graph- Comparison showing Average Latency

### 5.4. Inference

With a remarkable performance in terms of average latency and a significant increase in delivery probability we can surely say that affinity based movement model is preferable over generic RWP model i.e. there should be some nodes which have affinity towards the source or the destination. Epidemic routing with affinity based movement performs best in terms of message delivery probability and average latency.

## 6. APPLICATION SCENARIO

As an insight to practical application, we describe how the proposed concept can be utilized in VANETs. As mentioned earlier there have been a number of significant applications of opportunistic networks like ZebraNet. Consider a simple application where opportunistic networks can be deployed for measuring and monitoring road traffics [23]. Suppose we have a number of traffic monitoring/data collecting devices (represented by *source* in fig. 10), placed at important junctions, which collect traffic data like queue length, average waiting time and average vehicle velocity. A number of probe vehicles equipped with short range radio devices (represented as *R-readers*) are deployed which act as satellite nodes and help in relaying data from the source to the destination (or gateway). The destination may be a central monitoring station or a data collecting hub or another vehicle. These probe vehicles move along traffic flow and their movement are confined in and around the source. Data is relayed from one reader to the other until they are finally delivered to the destination where they may be processed and analysed for studying traffic flow at each junction. Moreover, on road vehicles can be informed about congestion ahead or expected travel time (to a given destination) by routing data *opportunistically* to these vehicles. Such application can be easily deployable, scalable and have a very low maintenance cost. Such networks are highly appreciated since data can be delivered opportunistically with minimum infrastructure and with no direct path between the source and the destination.

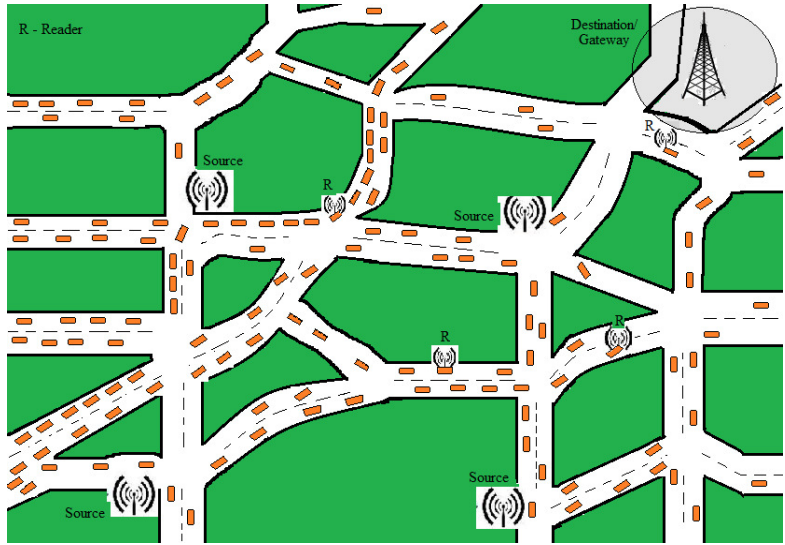

Figure 10. A simple application for measuring and monitoring traffic flow.

## 7. CONCLUSION

This paper describes a simple opportunity based message delivery protocol between two distantly placed source and destination nodes. We first provide an estimate of some of the important factors/parameters like number of encounters, contact or link duration and inter-contact time which affect the performance of opportunistic network. We find that the above factors are directly related to node density, node velocity and movement pattern of the nodes for a given simulation area. Hence, we propose a naive affinity based mobility model.

We deploy a number of *helper and satellite nodes* which carry out the task of coordinating the communication between two distantly placed source and destination. Simulation results for affinity based mobility model shows that there is significant increase in throughput and decrease in packet delivery latency while having the same overhead ratio when applied to all the three routing protocols viz. epidemic, spray-and-wait and prophet.

Though we have investigated some of the important aspects of opportunistic networks and demonstrated how network performance can be improved with affinity based mobility model, there are still some open areas which needs further research. Increase in node density (with respect to the given scenario) for a given simulation area has a significant impact on the network throughput. This paper does not include the effect of node density on the throughput and delivery latency which can another vital area of research. Moreover, the question that still needs to be answered are – what is the optimal node density having pre-specified node characteristics for the given scenario and how changing the degree of bias affects network performance?

ACKNOWLEDGEMENTS

The authors would like to thank all the faculty members of department of information technology, Jadavpur University for their support and encouragement. We are also grateful towards DST, Govt. of India for supporting our work under the PURSE project.

**APPENDIX**

## A. Number of encounters

From eq. (i) we know that probability distribution of a node, having RWP movement, with the area *'D'* is given by:

$$p = P(x, y \in D) = \iint_D f_{X,Y}(x, y) \, dx \, dy, \text{ where}$$

$$f_{X,Y}(x, y) \approx \frac{36}{a^6}\left(x^2 - \frac{a^2}{4}\right)\left(y^2 - \frac{a^2}{4}\right), for\ x \in \left[-\frac{a}{2}, \frac{a}{2}\right] and\ y \in \left[-\frac{a}{2}, \frac{a}{2}\right]$$

$$hence,\ \ p = \frac{36}{A^3} \iint_D \left(x^2 - \frac{a^2}{4}\right)\left(y^2 - \frac{a^2}{4}\right) dxdy,\ where\ A = a^2$$

$$\frac{dp}{dA} = -\frac{3p}{A} \quad \ldots (v)$$

Total time spent with area *D* is equal to number of encounters with the static node, with each encounter being of duration (*d*) E[$T_{contact}$] [eq. (iii)]. Hence,

$$N_{encounter} \times d = p \times T_{sim}$$

$$or, \; dN = \frac{dp \times T_{sim}}{d} \quad \ldots \text{(vi)}$$

$$or, \; dN = -\frac{3N}{A} dA \quad \text{[substituting (v) in (vi)]}$$

Integrating both sides, $\int \frac{dN}{N} = -3 \int \frac{dA}{A}$, we get

$$or, \; N_{encounter} = \frac{c}{A^3}$$

Hence, we note that number of encounters is inversely proportional to the cube of simulation area.

## B. Prophet router initialization constants

PRoPHET [22] is the Probabilistic Routing Protocol using History of past Encounters and Transitivity, which is used to estimate each node's delivery probability for each other node. When node i meets node j, the delivery probability of node i for j is updated by

$$p'_{ij} = (1 - p_{ij})p_0 + p_{ij}$$

where $p_0$ is the initial probability.

When node *i* does not meet *j* for some time, the delivery probability decreases by $p'_{ij} = \alpha^k \times p_{ij}$ where α is the aging factor and k is time units since last update.

The PRoPHET protocol exchanges index messages as well as delivery probabilities. When node i receives node j's delivery probabilities, node i may compute the transitive delivery probability through j to z with

$$p'_{iz} = p_{iz} + (1 - p_{iz})p_{ij}p_{jz}\beta$$

where β is the design parameter for the impact of transitivity.

## C. Spray-And-Wait Protocol

The Spray and Wait protocol [21] creates a number of copies *N* to be transmitted (*sprayed*) per message. In its normal mode, a source node *A* forwards the *N* copies to the first *M* different nodes encountered. In the binary mode, any node *A* that has more than one message copies and encounters any other node *B* that does not have a copy, forwards to *B* the number of *N/2* message copies, and keeps the rest of the messages. A node with one copy left only forwards it to the final destination.

**Authors**

Suvadip Batabyal is a Senior Research Fellow in department of Information Technology, Jadavpur University. He received his B.E. degree in 2007 and M.E. degree in 2011. Earlier he worked as systems engineer in Tata Consultancy Service Ltd for a period of 2 years. His current areas of interest are software and systems researches on communication protocols and applications for mobile, ad hoc and  sensor networks. Special research directed towards emerging wireless concepts like Delay Tolerant Networks (DTNs) and Opportunistic networks (OpNets), applications of DTNs in Vehicular Adhoc Networks (VANET) and road traffic analysis with respect to Indian scenario.

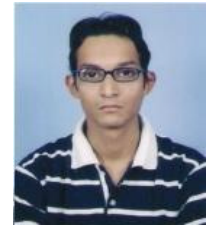

Dr. Parama Bhaumik is a PhD in Engineering from Jadavpur University (2009), and M.Tech & B.Tech in Computer Science & Engineering from Calcutta University, India (2002). She is currently working as Assistant Professor in Jadavpur University. She has more than 15 research publications in international conferences and journals of repute. Her present research activities include: Wireless sensor network, Mesh Networking, Opportunistic Networks and Topology Management using intelligent methods.

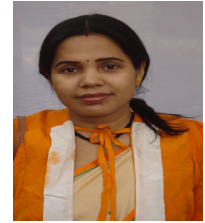